\documentclass[fleqn,twoside]{article}
\usepackage[headings]{espcrc2}
\usepackage{amssymb}
\readRCS
$Id: espcrc2.tex,v 1.2 2004/02/24 11:22:11 spepping Exp $
\usepackage[figuresright]{rotating}

\newcommand{\be}{\begin{equation}}
\newcommand{\ee}{\end{equation}}
\newcommand{\bea}{\begin{eqnarray}}
\newcommand{\eea}{\end{eqnarray}}
\newcommand{\beas}{\begin{eqnarray*}}
\newcommand{\eeas}{\end{eqnarray*}}

\newtheorem{theorem}{Theorem}

\def\build#1_#2^#3{\mathrel{\mathop{\kern 0pt#1}\limits_{#2}^{#3}}}

\def\One{\mathbb{I}}
\newcommand{\AmS}{{\protect\the\textfont2
  A\kern-.1667em\lower.5ex\hbox{M}\kern-.125emS}}

\title{An \'Etude in non-linear Dyson--Schwinger Equations\thanks{work supported in
parts by grant NSF-DMS0401262 at Center for Mathematical Physics, Boston University}}

\author{Dirk Kreimer\address{Institut des Hautes \'Etudes Scientifiques,\\
        91440 Bures-sur-Yvette, France}\thanks{supported by CNRS}\\
        Karen Yeats\address{Math.\ Dept., Boston Univ.,\\  Boston, MA02215, USA}
        }
\runtitle{Non-linear Dyson--Schwinger equations} \runauthor{D.\
Kreimer and K.\ Yeats}

\begin{document}

\begin{abstract}
We show how to use the Hopf algebra structure of quantum field
theory to derive nonperturbative results for the short-distance
singular sector of a renormalizable quantum field theory in a simple
but generic example. We discuss renormalized Green functions
$G_R(\alpha,L)$ in such circumstances which depend on a single scale
$L=\ln q^2/\mu^2$ and start from an expansion in the scale
$G_R(\alpha,L)=1+\sum_k \gamma_k(\alpha)L^k$. We derive recursion
relations between the $\gamma_k$ which make full use of the
renormalization group. We then show how to determine the Green
function by the use of a Mellin transform on suitable integral
kernels. We exhibit our approach in an example for which we find a
functional equation relating weak and strong coupling expansions.
\vspace{1pc}
\end{abstract}

\maketitle

\section{The structure of Green functions}
Our interest is the high energy sector of a renormalizable quantum
field theory. We want to explore consequences of the underlying Hopf
algebra structure to obtain non-perturbative results, using the
self-similarity of Green functions.

Our approach is based on the Dyson--Schwinger equations for
one-particle irreducible renormalized Green functions. As we want to
obtain non-perturbative results the choice of a renormalization
condition for us means simply the choice of a boundary condition for
the full Green function.

In this short paper we want to exhibit the basic idea underlying
such a program. We focus on the question how to treat the
non-linearity of Dyson--Schwinger equations systematically. We
assume we work in a renormalizable quantum field theory which
provides a finite set ${\cal R}$  of amplitudes which need
renormalization.

For a given superficially divergent amplitude $r\in {\cal R}$ we let
$\Gamma^r$ be the sum \be \Gamma^r=\One+\sum_{\Gamma} \alpha^{|\Gamma|}
\frac{\Gamma}{{\rm sym}(\Gamma)}\label{dsesum}\ee over all 1PI
graphs $\Gamma$ contributing to that amplitude, with $\alpha$ a
loop-counting small parameter.  Projection onto suitable form
factors $\phi(r)$ allows the sum to start with one, so that by
the application of the Feynman rules $\phi(\Gamma^r)$ is the
corresponding structure function and the Lagrangian $L$ is given by
\be L=\sum_{r\in{\cal R}}\phi(r).\ee

We can then write \cite{anatomy} \be \Gamma^r =
\One+B_+(\Gamma^r,Q(\{\Gamma^i\})),\label{primsum},\ee where
$Q=Q(\{\Gamma^i\}_{r\in {\cal R}})$ evaluates to an invariant charge
under the Feynman rules and the Hochschild one-cocycle \be
B_+^r(\Gamma^r,Q)=\sum_{k\geq 1}\alpha^k B_+^{k;r}(\Gamma^r Q^k)\ee
is a sum of one-cocycles $B_+^{k,r}$ and $Q$ is a monomial in the
$\Gamma^r$. The uniqueness of $Q$ implies the Slavnov--Taylor
identities for the renormalized couplings \cite{anatomy}.

The $B_+^{k,r}$ themselves are obtained from the skeleton graphs
$\gamma$ of the theory: \be B_+^{k,r}=\sum_\gamma \frac{1}{{\rm
sym}(\gamma)}B_+^{\gamma},\ee where the sum is over all Hopf algebra
primitives $\gamma$ contributing to the amplitude $r$ at $k$ loops.
These maps are defined to be closed Hochschild one-cocycles on the
sub Hopf algebra generated by their concatentations and products
\cite{anatomy,BergbKr}.

Effectively, (\ref{primsum}) reduces the sum (\ref{dsesum}) over all
graphs to a sum over primitive ones, making use of the recursive
structure of this fixpoint equation which determines the sum of
graphs which contribute to  a chosen amplitude. The sums involved
typically reflect the universal law of \cite{Y} and will be
discussed in detail in upcoming work.

We set \be \Gamma^r= \One+\sum_j c_j^r\alpha^j\ee and those $c_j^r$ are
the linear generators of a sub-Hopf algebra:
\begin{theorem}
There exists maps $B^{k;r}_+$, polynomials $P^r_{k,j}$ in those
linear generators  and integers $s_r$ such that \bea \Gamma^r & = &
\One+\sum_k \alpha^k
B^{k;r}_+\left(\Gamma^r Q^{k}\right),\\
\Delta B_+^{k;r} & = &  B_+^{k;r}\otimes \One + \left({\rm
id}\otimes B_+^{k;r}\right)\Delta,\\
Q & = & \alpha \prod_{r\in{\cal R}} [\Gamma^r(\alpha)]^{s_r},\\
\Delta c_k^r & = & \sum_{j=0}^k P_{k,j}^r\otimes c^r_{k-j},\eea
which make the system $\{c^r_k\}$ into a sub Hopf algebra
$H(\Delta,m,S,\epsilon)$ of the Feynman graph Hopf algebra.
\end{theorem}
The polynomials $P^r_{k,j}$ are easily determined and we refer the
reader to \cite{anatomy} for details.

 Feynman rules are then defined in accordance with the Hochschild
cohomology: \be \phi\left(B_+^\gamma(h)\right)(\{q\})=\int\!\!
d_\gamma(\{k\},\!\{q\}) \phi\!\left(h\right)\!(\{k\}).\label{FR}\ee
Here $ d_\gamma(\{k\},\{q\})$ is a measure determined by the
primitive $\gamma$ which depends on internal loop momenta $\{k\}$
and external momenta $\{q\}$ such that the Hopf algebra primitive
determines the kernel $d_\gamma$ and hence the Feynman rules by \be
\phi(B_+^\gamma(\One))=\int d_\gamma(\{k\},\{q\}) ,\ee and an
appropriate insertion of $\phi(X)(\{k\})$ in (\ref{FR}) into the
integrand provided by $d_\gamma$ in accordance with the pre-Lie
structure of graphs is understood. Similarly we define renormalized
Feynman rules by iterating in a subtracted kernel \be
\phi_R\left(B_+^\gamma(\One)\right)=\int\!\!
\left[d_\gamma(\{k\},\!\{q\})\!
-d_\gamma(\{k\},\!\{\mu\})\right],\ee where $\{\mu\}$ indicates a
suitable renormalization point.

There exists a basis of graphs and external structures in the Hopf
algebra such that \be \phi_R(Q)=\phi_R(Q)(L),\ee where $L=\ln
q^2/\mu^2$ is a single scale which determines the running of the
invariant charge. The choice of such a basis disentangles internal
subdivergences into divergent contributions which depend on a single
scale and finite contributions which determine the set of primitive
elements in such a base. In this base, short distance singularities
are captured by Green functions which are functions of two
dimensionless variables $\alpha,L$, with a remarkable duality
between these two variables first observed in \cite{BrK}.

In perturbation theory the Feynman rules now allow us to write \be
G_R^r(\alpha,L)=\phi_R(\Gamma^r)=1+\sum_k \alpha^k\phi_R(c^r_k)(L).
\ee We can expand in a different manner \be G_R^r(\alpha,L)=1+\sum_k
\gamma_k^r(\alpha)L^k,\label{Lexp}\ee and the renormalization group
dictates relations between the $\gamma_k^r$. We work them out in a
moment.

First, we note that in the case of a linear DSE \cite{BergbKr}, we
get \be
\partial_L \phi(Q)(L)=0,\ee and hence a scaling solution \be
G(\alpha,L)=e^{-\gamma(\alpha)L}\ee solves the linear DSE  so that
\be \gamma_k(\alpha)=\frac{\gamma_1(\alpha)^k}{k!}.\ee

To proceed in general we consider the map \be P^{(n)}_{\rm
lin}=\underbrace{P_{\rm lin}\otimes\cdots\otimes P_{\rm lin}}_{n\;
\rm times}\Delta^{n-1}\ee where $P_{\rm lin}$ is the projector into
the linear span of generators of the Hopf algebra. From
\cite{anatomy,BergbKr} we have: \begin{theorem} The linearized
coproduct is obtained as $$  P_{\rm lin}^{(2)}\Gamma^r = P_{\rm
lin}\Gamma^r\otimes P_{\rm lin}\Gamma^r+P_{\rm lin}Q\otimes
\alpha\partial_\alpha \Gamma^r,$$ where \be P_{\rm lin}Q=\sum_r s_r
P_{\rm lin}\Gamma^r.\ee\end{theorem} This allows us to understand the
iterative structure of the next-to$\ldots$ leading log expansion
(\ref{Lexp}).

We define for $n>1$ \be
\sigma_n:=\frac{1}{n!}\;m^{n-1}\;\underbrace{\sigma_1\otimes\cdots
\otimes \sigma_1}_{n\;{\rm times}}\;\Delta^{n-1},\label{sig}\ee and
$\sigma_1$ is the residue defined by \be
\sigma_1=\partial_L\phi_R\left(S\star Y_{\rm
aug}\right)(L)|_{L=0}.\ee

Actually, $\sigma_n$ evaluates to the coefficient of the $L^n$ term
in the evaluation of a Hopf algebra element by the renormalized
Feynman rules, by the scattering type formula \cite{CK}.

We have \be h\not\in H_{\rm lin}\Rightarrow \sigma_1(h)=0,\ee so we
can use Theorem 2 and, by the above definition (\ref{Lexp}) of
$\gamma_k^r(\alpha)$, \be
\gamma_k^r(\alpha)=\sigma_k(\Gamma^r(\alpha)).\ee

Projection onto the linear generators delivers the desired formula
for the expansion in $L$: \bea
\gamma_k^r(\alpha) & = & \frac{1}{k}\left[\gamma_1^r(\alpha)\gamma_{k-1}^r(\alpha)\right.\nonumber\\
 & & \left. +\sum_j
s^j\gamma_1^j(\alpha)\alpha\partial_\alpha\gamma_{k-1}^r(\alpha)\right].\label{nextto}\eea

With the the above choice of basis we can now introduce the Mellin
transform \be F_\gamma(\rho)=\int \phi(B_+^\gamma(\One))
[k^2]^{-\rho}\ee (with obvious generalizations to the multivariate
case as studied below for example) and the DSE turns into an
equation which determines $\gamma_1^r$ as we will see in  a moment,
while the further terms in the $L$ expansion are determined from
(\ref{nextto}) above.

The Green function also has the usual expansion in $\alpha$ which is
triangular wrt $\gamma_k$ \be \gamma_k^r(\alpha)=\sum_{j\geq
k}\gamma_{k,j}^r\alpha^j.\ee We can hence proceed to work out the
recursion relations which express the functions $\gamma_k^r$ through
the functions $\gamma_1^r$ for $k>1$, and turn the Dyson--Schwinger
equations into an implicit equation which allows to determine  the
sole unknown functions $\gamma_1^r(\alpha)$ from the knowledge of
the above Mellin transforms. A full discussion is given in future
work. We now exhibit the approach in an example.

\section{A simple example}
For concreteness, we consider massless Yukawa theory and consider
all self-iterations of the one-loop massless fermion propagator,
with subtractions in the momentum scheme at $q^2=\mu^2$. Our Green
function is an inverse propagator with momentum $q$ and a function
of two variables $a$ and $L=\ln q^2/\mu^2$. We ignore radiative
corrections at the bosonic line and also at vertices, so the set
${\cal R}$ has a single element and the superscript $r$ is
suppressed henceforth. We rederive the results of \cite{BrK} for
this case.

We write the perturbative series for the Dyson--Schwinger equation
as \be X(a)=\One-a B_+^c\left(\frac{1}{X(a)}\right),\ee where $\int
\phi(B_+^c(\One))$ provides the one-loop self-energy integral to be
iterated. Note that upon setting $X(a)=\One-\underline{X}(a)$, this
is the equation for the self-energy $\underline{X}(a)=-P_{\rm
lin}X(a)$ studied in \cite{BrK}.

With \be Q=1/X^2\rightarrow P_{\rm lin}(Q)=-2\underline{X}(a),\ee we
find the linearized coproduct \be P^{(2)}_{\rm lin}X(a) = P_{\rm
lin}X(a)\otimes (P_{\rm lin}-2a\partial_a)X(a).\ee This is
Proposition 1 of \cite{BrK} and we also get
\begin{theorem}
The next-to next-to$\ldots$ leading log expansion in $L$ is given
through the anomalous dimension $\gamma_1(a)$ as \be
\gamma_k(a)=\frac{1}{k}\gamma_1(a)(1-2a\partial_a)\gamma_{k-1}(a).\ee
\end{theorem}
This is Proposition 2 of \cite{BrK}.

We can now work out with ease the recursions which express
$\gamma_k$, $k>1$ through the Taylor coefficients of $\gamma_1$, for
example \bea \gamma_2 & = &
\frac{1}{2}\left(\gamma_1^2-2\gamma_1a\partial_a\gamma_1
 \right)\\
 & = & \frac{1}{2}\left[-\gamma_{1,1}^2a^2-4\gamma_{1,1}\gamma_{1,2}\alpha^3+\cdots
\right],\\
\gamma_3 & = &
\frac{1}{6}\left(\gamma_1(1-2a\partial_a)\gamma_1(1-2a\partial_a)\gamma_1\right)\\
 & = & \frac{1}{6}\left( 3a^3\gamma_{1,1}^3+\cdots\right),\eea
 and so on.

Such recursions are obtained for any non-linear DSE by iterating
Theorem 2. Also, we observe that we actually only need the
cocommutative part in the determination of the coproduct as is
evident from the very cocommutative definition (\ref{sig}) of
$\sigma_k$, $k>1$. The non-cocommutative part is always of lower
degree in $L$ in the obvious filtration by $L$.

 It remains to understand how to
compute $\gamma_1(\alpha)$. Instead of an explicit analysis of
non-linear ODEs as in \cite{BrK} we proceed here by the Mellin
transform, as promised. Here, it reads \bea F(\rho) & = &
\frac{1}{q^2}\int d^4 k
\frac{k\cdot q}{[k^2]^{1+\rho}(k+q)^2}_{|_{q^2=1}}\nonumber\\
& =: & \frac{r}{\rho}+\sum_{i\geq 0}f_{i}\rho^i. \eea In our simple
example we have \be
F(\rho)=\frac{-1}{\rho(2-\rho)}.\label{Mellin}\ee

Let us introduce a short hand notation: \be \gamma\cdot
U=\sum_{k=1}^\infty \gamma_k(\alpha) U^k.\ee

Then, the Dyson--Schwinger equation becomes \be \gamma\cdot L=\alpha
(1+\gamma\cdot
\partial_{-\rho})^{-1}[e^{-L\rho}-1]F(\rho)|_{\rho=0},\label{dsem2}\ee where we evaluate the
rhs at $\rho=0$ after taking derivatives. The functional dependence
of the non-linear DSE on $XQ=X^{-1}$ reflects itself on the rhs.

The only unknown quantities in this equation are the Taylor
coefficients $\gamma_{1,j}$ which are implicitly defined through the
Taylor coefficients of the Mellin transform (\ref{Mellin}) above.

Taking a derivative of (\ref{dsem2}) wrt $L$ and setting $L$ to zero
allows us to read them off: \bea \gamma_1
 & = & \alpha(1+\gamma\cdot \partial_{-\rho})^{-1} \rho F(\rho)|_{\rho=0}\\
 & = & \alpha r +\alpha
\left(\sum_{k\geq
1}[\gamma\cdot\partial_{-\rho}]^k\right)\times\nonumber\\
 & & \times\left[\sum_{k=1}^\infty
\rho^k f_{k-1}\right]|_{\rho=0},\label{DSErec}\eea so
$\gamma_{1,1}=r$ universally.

In our example $Q=X^{-2}$ we furthermore find \bea
\gamma_{1,2} & = & rf_0,\\
\gamma_{1,3} & = & rf_0^2+r^2 f_1,\eea and so on.

In this manner one confirms the results of \cite{BrK} which serve
here as a mere example for a much more general approach. Note that
working with the toy Mellin transform $r=f_i=1$ reproduces the
generating functions which counts the graphs contributing at each
loop order confirming the count of Wick contractions in \cite{BrK2},
in terms of Catalan numbers $$1,1,2,5,14,42,\cdots.$$ For the
anomalous dimension itself we indeed also confirm the series A000699
of \cite{EncInt}, \be {\rm A}000699=1,1,4,27,248,2830,\cdots,\ee
from Eq.(\ref{DSErec}) above, in accordance with \cite{BrK2}.

 Before we finish this paper by discussing a
double Mellin transform we mention one particular nice feature of
this example which we very much hope to work out in general in the
future.

\section{A functional equation}
It is our hope that eventually a non-perturbative renormalized Green
function can be related to suitably defined $\zeta$-functions. The
existence of a combinatorial Euler product underlying the
decomposition of graphs into primitive graphs is a first hint
\cite{Kr}.

Here we report on another such hint based on the possibility to
reformulate the result of \cite{BrK} such that a functional equation
is obtained. Inspection of the solution in \cite{BrK}  (which is
also suggested by the functional equation of the complementary error
function) shows \bea
  \widetilde{\Sigma}(a, p)
  & = &  - \frac{\sqrt{a/(2\pi)}}{\exp(p^2){\rm erfc}(p)}\times\nonumber\\
   & & \times \widetilde{\Sigma}\left(\frac{(\exp(p^2){\rm erfc}(p))^4}{a/(2\pi)^2},
  p\right),\label{zetaf}
\eea where $a$ is now the loop counting parameter and  $p$ is
another variable such that with $z=e^{2L}$, \be
p=\frac{d}{dz}\sqrt{\frac{2}{a}}\left(z-z\widetilde{\Sigma}\left(\mu^2\sqrt{z}\right)\right).
\ee
 Note that on the lhs of (\ref{zetaf}) we have a weak
coupling expansion for $a$, on the rhs we have a strong coupling
expansion, hence an expansion in $1/a$.

With $T = a/(2\pi)$, $u = (\exp(p^2){\rm erfc}(p))^{-4}$, and $Z(T,
u) = \widetilde{\Sigma}(a, p)$ we get a functional equation
reminiscent of a functional equation for a $\zeta$-function in two
variables for the function field case \cite{Deninger,Pell} for the
non-perturbative renormalized Green function \be
  Z(T, u) = -u^{\frac{5}{4} - 1}T^{2(\frac{5}{4}-1)}Z\left(\frac{1}{Tu},
  u\right).
\ee The propagator coupling duality of this Green function can now
be expressed with  $u=\exp(s+t)$, $T=\exp(-t)$, and $\zeta(s, t) =
\exp(\frac{t-s}{8})Z(T,u)$  as
\be
  \zeta(s, t) = - \zeta(t, s),
\ee which we report here for motivation to think further about the
connection between quantum field theory and $\zeta$-functions.
\section{The appearance of transcendentals}
How do we continue in the case where we have several insertion
places? Nothing changes in the above derivation apart from the fact
that we now have to work with our double Mellin transform due to the
fact that now both propagators obtain logarithmic corrections. As
the primitive self-energy $B_+^c(\One)$ now obtains corrections at
the internal fermionic and bosonic line, we are led to consider a
coupled system including also the bosonic propagator.

If we are hence to consider the system (based now on two elements in
${\cal R}$, fermion and boson self-energies) \bea X(a) & = & \One
-aB^c_+\left(\frac{1}{X(a)Y(a)}\right)\nonumber\\ & = & \One
-aB^c_+\left(X(a)Q(a)\right),\\
Y(a) & = & \One-B^b_+\left(\frac{1}{X(a)^2}\right)\nonumber\\ & = &
\One -aB^b_+\left(Y(a)Q(a)\right), \eea with $Q^{-1}(a)=X^2(a)Y(a)$
and \be P_{\rm lin}Q(a)=-2P_{\rm lin}X(a)-P_{\rm
lin}Y(a),\label{n2}\ee describing all possible iterations of
massless fermion and scalar one-loop graphs, the corresponding
functions $\gamma^X_k(a),\gamma^Y_k(a)$ are determined in terms of
the anomalous dimensions $\gamma^X_1(a),\gamma^Y_1(a)$. The latter
are obtained from the system of Dyson--Schwinger equations \bea
\gamma_1^X & = & \alpha (1+\gamma^X\cdot
\partial_{-\rho_1})^{-1} (1+\gamma^Y\cdot
\partial_{-\rho_2})^{-1}\times \nonumber\\
 & & \times
 (\rho_1+\rho_2)F_a(\rho_1,\rho_2)|_{\rho_i=0},\label{dse1}\\
\gamma^Y_1 & = & \alpha (1+\gamma^X\cdot
\partial_{-\rho_1})^{-1} (1+\gamma^X\cdot
\partial_{-\rho_2})^{-1}\times \nonumber\\
 & & \times
 (\rho_1+\rho_2)F_b(\rho_1,\rho_2)|_{\rho_i=0}.
 \label{dsem}\eea
 We thus have to consider two Mellin transforms in
two variables each. They read for the fermion self-energy
$B_+^c(\One)$ \bea F_c(\rho_1,\rho_2) = \frac{1}{q^2}\int d^4k
\frac{k\cdot q}{[(k+q)^2]^{1+\rho_1}[k^2]^{1+\rho_2}}\nonumber\\
  =
\frac{-1+\rho_2}{(2-\rho_1-\rho_2)}
\frac{e^{\sum_{k=1}^\infty-2\zeta(2k+1)f_k(\rho_1,\rho_2)}}{(1-\rho_1-\rho_2)(\rho_1+\rho_2)},\eea
and for the boson self-energy $B_+^b(\One)$ \bea F_b(\rho_1,\rho_2)
= \frac{1}{q^2}\int d^4k
\frac{k\cdot (k+q)}{[(k+q)^2]^{1+\rho_1}[k^2]^{1+\rho_2}}\nonumber\\
  =
\frac{1-(\rho_1+\rho_2)+\rho_1\rho_2}{(2-\rho_1-\rho_2)}\times\nonumber\\
\times
\frac{e^{\sum_{k=1}^\infty-2\zeta(2k+1)f_k(\rho_1,\rho_2)}}{(1-\rho_1-\rho_2)^2(\rho_1+\rho_2)},\label{mel2}\eea
 where traces have been taken to obtain the relevant structure functions and where
 $f_k(\rho_1,\rho_2)$ are the two-variable symmetric polynomials given by
\be
f_k(\rho_1,\rho_2)=\sum_{j=1}^{2k}\frac{2k!}{j!(2k+1-j)!}\rho_1^j\rho_2^{2k+1-j}.\ee
 We
observe that the appearance of transcendentals is utterly generated
from the presence of a second insertion place.

Eqs.(\ref{dse1}--\ref{mel2}) completely determine the anomalous
dimensions in question and together with (\ref{nextto},\ref{n2}) the
two Green functions. To what extent this leads to further functional
equations is under current investigation.
\section*{Acknowledgment} Both authors thank David Broadhurst for
discussions and proofreading the ms.

\end{document}